\documentclass[12pt,preprint]{aastex}

\shorttitle{On the Moon's Distance and Motion}
\shortauthors{Krisciunas} 

\begin{document}
\received{8 November 2013}

\title{A Mixture of Ancient and Modern Understanding Concerning 
the Distance and Motion of the Moon}
\author{
Kevin Krisciunas\altaffilmark{*}
}
\altaffiltext{*}{George P. and Cynthia Woods Mitchell Institute for Fundamental 
Physics \& Astronomy, Texas A. \& M. University, Department of Physics,
  4242 TAMU, College Station, TX 77843; {krisciunas@physics.tamu.edu} }

\begin{abstract} 

Ptolemy's model of the Moon's motion implied that its distance
varies by nearly a factor of two, implying that its angular
size should also vary by nearly a factor of two.
We present an analysis of 100 naked eye observations of the
Moon's angular size obtained over 1145 days, showing regular variations
of at least 3$^{\prime}$.  Thus, ancient
astronomers could have shown that a key implication of Ptolemy's
model was wrong.  In modern times we attribute the variation of
distance of the Moon to the combined effect of the
ellipticity of the Moon's orbit and the
perturbing effect of the Sun on the Earth-Moon system.
We show graphically how this affects the ecliptic longitudes
and radial distance of the Moon.  The longitude and distance ``anomalies''
are correlated with the Moon's phase.  This is
illustrated without any complex equations or geometry.

\end{abstract}


\section{The Angular Size of the Moon}

We have known since the time of Aristarchus (3rd century B.C.) that the
Moon is about 60 Earth radii distant, and its angular diameter is roughly
half a degree.  The mean angular size is actually 31.1 arc minutes.

In the {\em Almagest} Ptolemy gives the correct {\em range} of the angular
size of the Moon, about 4 arc minutes.  His values of the 
minimum and maximum angular size are about 2 arc minutes too large.
The ancient Babylonians
and Greeks were more concerned with the {\em direction towards} the Moon,
not its physical distance.  Ptolemy's model of the motion of the Moon
implied that its distance ranged from 33.55 R$_{\oplus}$ (at first/third
quarter) to 64.17 R$_{\oplus}$ (at full/new Moon), nearly
a factor of two.\footnote[1]{See footnotes 7 and 8 of \citet{Kri09}.}
Since the angular size of the Moon in radians is just equal to its
diameter divided by its distance, Ptolemy's model implies that the
angular size of the Moon should also vary by nearly a factor of two.

Ptolemy's model was the most popular model of the Moon's motion until
the time of Tycho Brahe (1546-1601), which is curious, because even a casual
observer of the Moon would notice that the Moon's angular size does not vary by
a factor of two.  Even more curious is the scarcity of actual measurements
that have come down to us from ancient and medieval times.  To my
knowledge only Levi ben Gerson (1288-1344) and Ibn al-Shatir (1304-1375/6)
made such measures.  Levi claims to have 
measured the angular size of the Moon many times and found about the
same value at full/new Moon as at first/third quarter, in contradiction
to Ptolemy's model \citep{Gol85}.

An obvious question arises. Is it possible to measure regular variations of the
angular size of the Moon with the naked eye and derive the 
eccentricity of the Moon's orbit?
I fashioned a device that slides up and down a yardstick calibrated
in millimeters.  By eye I fit the Moon into a 6.2 mm hole.  Simple geometry
gives the angular size, with one proviso.  The pupil of the eye
is comparable in size to the hole for sighting the Moon.  
And the eye has a lens.  I was not
using a small ($\sim$1 mm) sight at the eye end.

Consider a 91 mm disk viewed at a distance of 10 meters.
It has the same angular size as the mean angular
size of the Moon.  I found that if I scale my raw angular sizes by
1.17 (from the measurements of a 91 mm disk), I obtain the
correct mean angular size of the Moon. 
Many of my 20 year-old 
students have obtained correction factors between 0.9 and 1.0,
while others get values as small as 0.7 or as large as 1.3.
An understanding of this is beyond the scope of the present paper.
Let us be content that my own scale factor eliminates a source
of systematic error related to my measurements.

From observations obtained in 2009 over 7 lunations, I was able to
demonstrate regular variations of the angular size of the Moon
\citep{Kri09}.  The uncertainty of my individual observations is
about 1 arc minute, which is accurate enough to show monthly
variations of 3-4 arc minutes.
Thus, it was within the capabilities of the ancient
Greeks not only to discover a serious problem with Ptolemy's model of the
motion of the Moon, but also to establish an approximately correct
value for the variation of angular diameter and distance.

Using 100 observations\footnote[2]{The raw data can be obtained
at: http://people.physics.tamu.edu/krisciunas/moon\_ang.html}
obtained over 1145 days (or 39 lunations), my
derived value of the {\em anomalistic month} (perigee to perigee
period) is 27.5042 $\pm$ 0.0334 days, which is 1.5 standard deviations
less than the official value of 27.55455 days.\footnote[3]{According
to \citet[][on p. 211]{Too81}, Hipparchus knew that the Moon returns
to the same velocity 4573 times in 126,007 days, 1 hour.  This
gives an anomalistic month of 27.55457 days.}

The measures of the Moon's angular size, folded by our derived value
of the length of the anomalistic month, are shown in Fig. \ref{phased}.  
Here I have used mathematical tools unavailable to pre-19th century astronomers,
namely a period finding algorithm \citep{Bre89} and a means of
estimating the uncertainty of the derived period \citep{Mon_OD99}.
My value of the eccentricity of the Moon's orbit is 0.039 $\pm$ 0.004,
which is noticeably smaller than the official modern value of 0.0549.

\section{Anomalies of the Moon's Motion}

What has been known about the Moon's motion, from ancient times to
the late 20th century, is summarized in a long and impressive
article by \citet{Gut98}.  On pp. 601-602 he briefly describes
the basics.  ``The Babylonians knew that the full moons could
be as much as 10 hours early or 10 hours late [compared to
uniform circular motion]; this is due to the
eccentricity $\epsilon$ of the Moon's orbit.''  This is known
as the first anomaly of the Moon's motion.  The Greeks would
have just considered it the largest epicycle of the Moon's orbit.

Gutzwiller continues, ``[T]he Greeks 
wanted to know whether the Moon displays the same kind of 
speedups and delays in the half moons, either waxing or waning.
The half moons can be as much as 15 hours early or 
late.  With the Moon moving at an average speed of slightly more than 
30$^{\prime}$ per hour ... it may be as much as 5 deg 
ahead or behind in the new/full moons; but in the half moons, it may be as much 
as 7 deg 30$^{\prime}$ ahead or behind its average motion.  This new feature is 
known as {\em evection}.'' This is the second anomaly of the Moon's motion.

The first anomaly causes deviations from the mean longitude up to 6 deg 
15$^{\prime}$.  The evection adds (sinusoidally) another 1 deg 15 $^{\prime}$.
Consider it the second largest epicycle.  
Tycho Brahe discovered the third largest perturbation of the longitudes,
known as the {\em variation}.  It has an amplitude of 40$^{\prime}$
and depends on twice the mean angular distance of the Moon from the Sun.

Tycho also discovered the ``annual inequality,'' which has an amplitude
of 11$^{\prime}$.  It accounts for the Moon slowing down in its motion
around the Earth when the Earth is near perihelion and speeding up
when the Earth is near aphelion.  Finally, Tycho discovered two
inequalities relating to the Moon's ecliptic latitudes.  In modern parlance
we would say that these anomalies are smaller and smaller
terms of a Fourier series that describes the Moon's motion.

Details of Tycho's lunar
theory are considerably beyond the scope of the present article.  
The reader is directed to the appropriate parts of Victor Thoren's 
excellent biography \citep{Tho90} and to \citet{Swe09}.  

How can we most easily understand this behavior of the Moon?  Consider
the gravitational force between the Sun and the Moon when
the Moon is 1 AU from the Sun.  Call this F$_{(\odot -Moon)}$.  It is equal
to 4.36 $\times 10^{20}$ Newtons.  Let the gravitational force between
the Earth and Moon when the Moon is at its mean distance of 384,400 km
be F$_{(\oplus -Moon)} \approx  1.98  \times 10^{20}$ Newtons.  The ratio
is $\approx$ 2.2.  In other words, the Sun-Moon gravitational force is
2.2 times stronger, on average, than the Earth-Moon gravitational force.
The Sun's effect on the Moon's motion must therefore be considerable!

In Fig. \ref{dang} we show the deviations of the Moon's motion from uniform 
circular motion during the year 2012, illustrating graphically what we have 
quoted from \citet{Gut98}.  The Moon's biggest deviations in longitude occur at 
first and third quarter.  The deviations at full and new Moon are 
smaller owing to the evection.

In Fig. \ref{r_2012} we plot the Earth-Moon distance vs. day of year for
2012.  We used values from the 2012 volume of the {\em Astronomical Almanac}.
Note that the apogee distance has a very small range from orbit
to orbit (63.3 to 63.7 R$_{\oplus}$), but the perigee distance ranges considerably,
from 56 to 58 R$_{\oplus}$.  If the Moon had an ideal, Keplerian elliptical
orbit, its distance would range from 5.5 percent closer than the mean value to
5.5 percent larger than the mean.  The effect of the Sun on the Moon's
orbit provides an unequal situation.  At maximum distance the Moon
is 5.8 percent further than the mean distance.  At minimum distance 
the Moon is 7.3 percent closer than the mean distance.  

In Fig. \ref{moon_rdiff} we provide a graph for each month of 2012,
showing the variation of distance of the Moon {\em with respect to the
Moon's mean ellipse} as a function of the difference of ecliptic
longitudes of the Sun and Moon.  In other words, in polar coordinates
we show the perturbation
in the radial direction due to the Sun's effect on the Moon's orbit.
During 2012 the Moon's orbit bulged out
by as much as 1.1 R$_{\oplus}$
at third quarter during the first three months.  Then it did so again
during July, August, and September, but at first quarter.  The radial
distance was smaller than the mean ellipse at full Moon by
1.0 R$_{\oplus}$ from March
through July, then was smaller by 0.9 to 1.0
R$_{\oplus}$ at new Moon from September through December.

\section {Discussion}

%The orientation of the orbit is not fixed in space, but precesses over time. 
%The nearest and farthest points in the orbit are the perigee and apogee 
%respectively. The line joining these two points (the line of apsides) rotates 
%slowly in the same direction as the Moon itself (direct motion), making one 
%complete revolution in 3232.6054 days or about 8.85 earth years.

We have demonstrated that it is possible to measure regular
variations of the Moon's angular size using naked eye observations.
The results shown in Fig. \ref{phased} are {\em consistent with}
Kepler's First Law, but we have not {\em proven} that the Moon's orbit is
an {\em ellipse}.  That would require much more accurate data.  In fact,
the Moon's path around the Earth is {\em not} even a closed orbit.  The {\em line
of apsides} (which connects the apogee and perigee) rotates slowly in
the same direction that Moon itself moves, with a period of
3232.6 days (about 8.85 years).  And the effect of the Sun
on the Moon's motion causes multiple anomalies in the ecliptic
longitudes and the radial distance that are correlated with the
phase of the Moon.

It should be pointed out that while the Babylonians and ancient
Greeks obtained positional measurements of the Moon that extended
over centuries, they did not take data in the ``modern'' way.
Starting with Tycho, astronomers have endeavored to identify
and eliminate sources of systematic error, and have quantified
and sought to reduce random errors.  

\citet{Neu75} pointed out, 
``In all ancient astronomy direct measurements and theoretical 
considerations are so inextricably intertwined that every correction
at any one point affects in the most complex fashion countless
other data, not to mention the ever present numerical inaccuracies and
arbitrary rounding which repeatedly have the same order of magnitude
as the effects under consideration.  In the history of the most
causal of all empirical sciences, in astronomy, the search for causes
is as fruitless as in all other historical disciplines.''

Because of the Moon's brightness and availability every month, it has been
an obvious candidate for study, from ancient to modern times.
Its motion is anything but simple.

%\vspace {1 cm}

\acknowledgments

We thank Don Carona for his ephemeris program that provided
machine readable values from the {\em Astronomical Almanac}
of the Moon's and Sun's ecliptic longitude, and the distance
to the Moon.

\figcaption[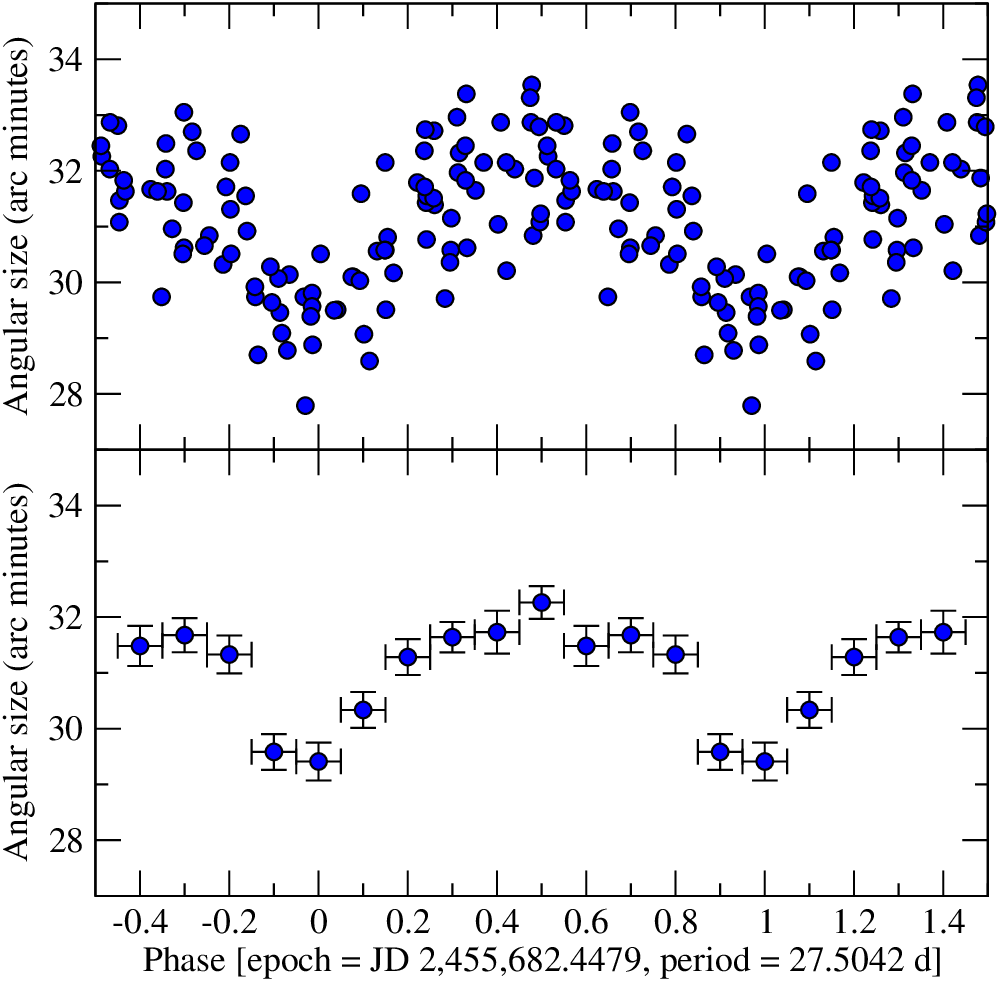]{Phased naked eye observations
of the angular size of the Moon.  Upper diagram: individual
data from April 21, 2009, through June 9, 2012.  Lower diagram:
averages for binned data. Modern period finding software
gives a period of 27.5042 $\pm$ 0.0334 days, which compares
well with the official modern value of 27.55455 days.
\label{phased}
}

\figcaption[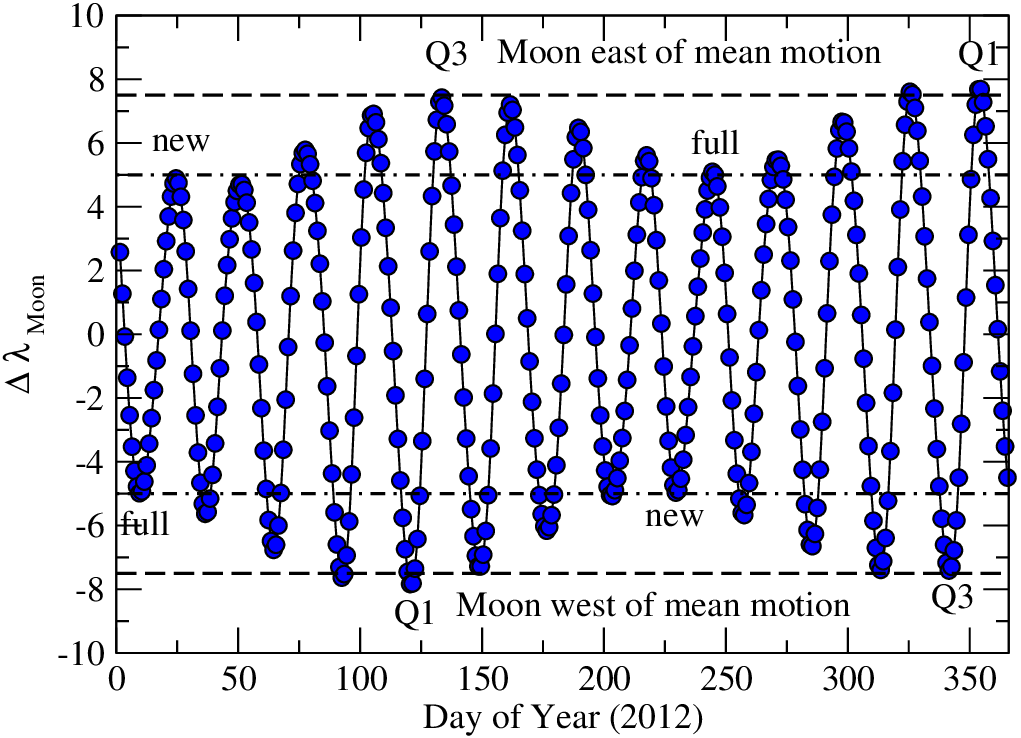]{Deviation of ecliptic longitude of Moon
(in degrees) compared to a perfectly circular orbit, for the year 2012.
We used information from the 2012 edition of the {\em Astronomical
Almanac}.
\label{dang}
}

\figcaption[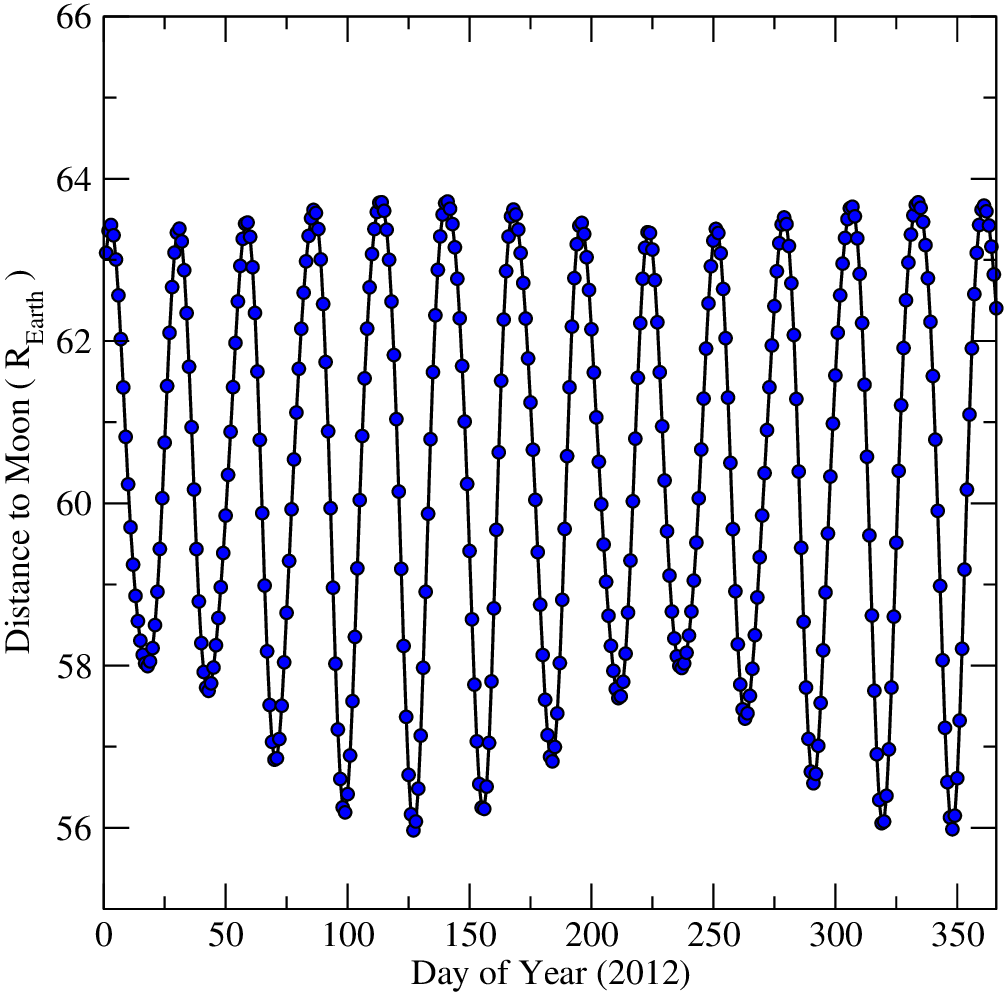]{Distance from the center of the Earth
to the center of the Moon, for each day of the year 2012.
Data are from the 2012 edition of the {\em Astronomical Almanac}.
\label{r_2012}
}

\figcaption[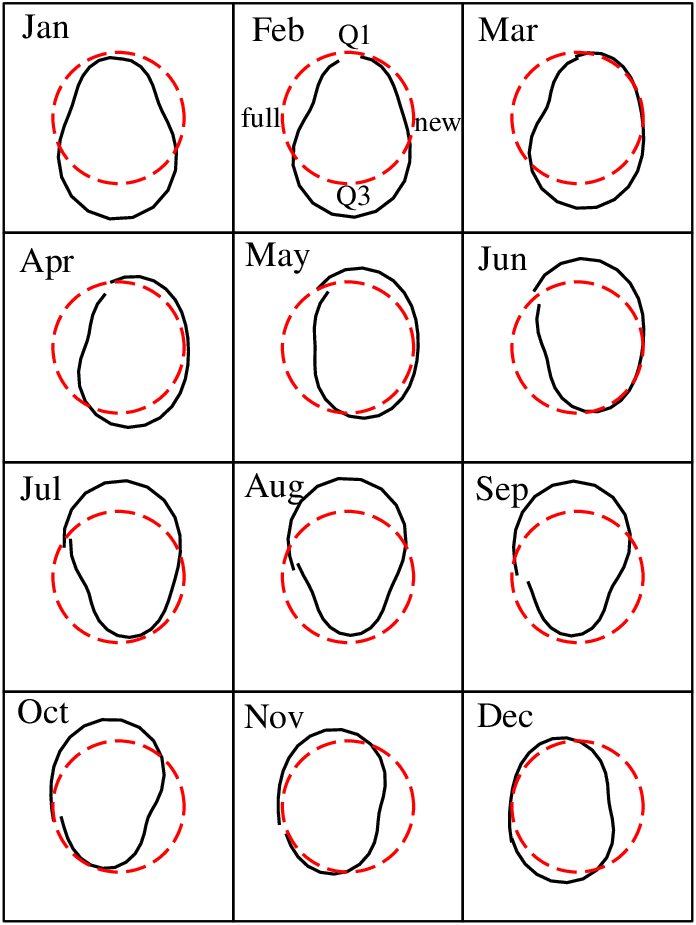]{Because of the Sun's
strong perturbation on the Moon's motion, the Moon's actual
distance from the Earth deviates from the basic (slowly rotating)
ellipse by up to 1.1 Earth radii.
We show this graphically in polar coordinates for the year 2012.  
In each of these graphs the Sun is far off the right hand side of the paper.
The new Moon, first quarter, full Moon, and third quarter are at the 
3 o'clock, 12 o'clock, 9 o'clock, and 6 o'clock positions, respectively.
The dashed circle has a radius of 2 R$_{\oplus}$, which gives
us a scale for the perturbations plotted in polar coordinates.
\label{moon_rdiff}
}

\clearpage

\begin{figure}
\plotone{phased.eps}
{\center Krisciunas Fig. \ref{phased}}
\end{figure}

\begin{figure}
\plotone{dang.eps}
{\center Krisciunas Fig. \ref{dang}}
\end{figure}

\begin{figure}
\plotone{r_2012.eps}
{\center Krisciunas Fig. \ref{r_2012}}
\end{figure}

\begin{figure}
\plotone{moon_rdiff.eps}
{\center Krisciunas Fig. \ref{moon_rdiff}}
\end{figure}

\end{document}